\documentclass[runningheads]{llncs}

\usepackage{url}
\usepackage{csquotes}
\usepackage{graphicx}
\usepackage{xcolor}
\usepackage{wrapfig}
\usepackage[hidelinks]{hyperref}
\usepackage{acro}

\usepackage[strings]{underscore}

\DeclareAcronym{gdpr}{
  short = GDPR,
  long  = General Data Protection Regulation
}
\DeclareAcronym{eu}{
  short = EU,
  long  = European Union
}
\DeclareAcronym{ap}{
  short = AP,
  long = access point
}
\DeclareAcronym{pii}{
  short = PII, 
  long = Personally Identifiable Information
}

\begin{document}
\title{mopri - An Analysis Framework for Unveiling Privacy Violations in Mobile Apps} 
\titlerunning{mopri - Privacy App Analysis}

\author{Cornell Ziepel$^{1}$ \and Stephan Escher$^{2}$ \and Sebastian Rehms$^{3}$ \and Stefan Köpsell$^{3}$}
\authorrunning{C. Ziepel et al.}
\institute{TU Dresden$^{1,3}$, SDTB$^{2}$, Germany\\
\email{research@cornell-ziepel.de$^{1}$, stephan.escher@sdtb.sachsen.de$^{2}$, \{firstname.lastname\}@tu-dresden.de$^{3}$}}

\maketitle              

\begin{abstract}
Everyday services of society increasingly rely on mobile applications, resulting in a conflicting situation between the possibility of participation on the one side and user privacy and digital freedom on the other. 
In order to protect users' rights to informational self-determination, regulatory approaches for the collection and processing of personal data have been developed, such as the EU's GDPR.
However, inspecting the compliance of mobile apps with privacy regulations remains difficult. 
Thus, in order to enable end users and enforcement bodies to verify and enforce data protection compliance, we propose mopri, a conceptual framework designed for analyzing the behavior of mobile apps through a comprehensive, adaptable, and user-centered approach.  
Recognizing the gaps in existing frameworks, mopri serves as a foundation for integrating various analysis tools into a streamlined, modular pipeline that employs static and dynamic analysis methods.
Building on this concept, a prototype has been developed which effectively extracts permissions and tracking libraries while employing robust methods for dynamic traffic recording and decryption. Additionally, it incorporates result enrichment and reporting features that enhance the clarity and usability of the analysis outcomes. The prototype showcases the feasibility of a holistic and modular approach to privacy analysis, emphasizing the importance of continuous adaptation to the evolving challenges presented by the mobile app ecosystem.

\end{abstract}

\section{Introduction}
Nowadays, mobile applications are an integral part of almost all areas of daily life.
Thereby, more and more services are exclusively accessible through such 
apps\footnote{\href{https://digitalcourage.de/blog/2021/digitalzwangmelder-analyse}{digitalcourage.de/blog/2021/digitalzwangmelder-analyse}, 
\href{https://bigbrotherawards.de/en/2023/deutsche-post-dhl}{bigbrotherawards.de/en/2023/deutsche-post-dhl}}. At the same time, integrated tracking services and the disproportionate collection and sharing of personal data and behavior are common practice in mobile apps \cite{kollnig_are_2022,razaghpanah2018apps}. 
This data is used in ways that can impact our reality \cite{slavkovik2021digital}.
On the basis of personalized profiles, content and in particular advertising is targeted at users \cite{ullah2023privacy}. For this purpose, collected and predicted data is also forwarded and sold to third parties, enabling, i. a., real-time auctioning of advertising spaces based on user's characteristics\footnote{\href{https://netzpolitik.org/2023/microsofts-datenmarktplatz-xandr-das-sind-650-000-kategorien-in-die-uns-die-online-werbeindustrie-einsortiert}{netzpolitik.org/2023/microsofts-datenmarktplatz-xandr-das-sind-650-000-kategorien-in-die-uns-die-online-werbeindustrie-einsortiert}}. In 2019 alone, the value of data sold in EU countries amounted to 75.3 billion euros\footnote{\href{https://www.iwd.de/artikel/der-datenmarkt-waechst-rasant-482082}{www.iwd.de/artikel/der-datenmarkt-waechst-rasant-482082}}, where this lucrative data-driven economy is also often named surveillance capitalism \cite{zuboff2023age}. 

This practice stands in stark contrast to individuals' \enquote{right to the protection of personal data}, as laid down in the EU's general data protection regulation (GDPR) \cite{gdpr}. The GDPR mandates that, in most cases, user consent is required for the collection of personal data. Additionally, users must be informed about the identity of the data controller and provided with clear insights into how their data will be processed and stored. However, inspecting the compliance of mobile apps with these privacy regulations remains difficult. 
Frequently, data collection is obscured by deceptive designs \cite{lupianez2022behavioural,nouwens2020dark} and, when present, described by overly complex, incomplete or even false consent banners and privacy policies \cite{andow_actions_2020,nguyen_share_2021}.

Consequently, there is a need for tools which enable comprehensive app analyses, to get a detailed understanding of the app's real data processing behavior and thus to enable end users and enforcement bodies to verify and enforce data protection compliance.
Due to common modular app development (using a variety of third-party libraries and services), such tools could also be beneficial for app developers themselves to get a clear understanding of their app behavior and operate it in compliance with the GDPR \cite{nguyen_share_2021}.
Currently, however, there is no easy-to-use tool available for the field of mobile applications; instead, there are a lot of components that can be used in complicated and time-consuming manual analyses setups. 
In addition, the complex and rapidly evolving mobile app ecosystem often causes approaches to quickly become obsolete. 

Therefore, we introduce mopri, an analysis framework for investigating the behavior of mobile apps through a comprehensive, adaptable, and user-centered approach. Recognizing the gaps of existing frameworks, mopri serves as a foundation for integrating various analysis tools into a streamlined, modular pipeline. It automates essential tasks, like installing and configuring necessary tools as well as recording and preparing analysis data. 
In addition to the investigation of the apps binary files, a key feature of mopri is its capability to execute the analyzed app in a controlled laboratory environment. This setup enables users to manually interact with the app while automatically recording and analyzing its network traffic, including transmissions that are encrypted during transport.
Moreover, collected data is enriched and visualized through interactive and static reports.
Such capabilities are crucial for promoting transparency and ensuring responsible privacy practices in mobile apps.
Building on this concept, a prototype of mopri has been developed specifically for analyzing
Android apps\footnote{\href{https://github.com/dfd-tud/mopri}{github.com/dfd-tud/mopri}}. The prototype showcases the feasibility of a holistic and modular approach to privacy analysis, emphasizing the importance of continuous adaptation to the evolving challenges presented by the mobile app ecosystem.

\section{Shining Light on App Behavior  - A Background} 
\label{sec:background}
In order to analyze an app with regard to its privacy-sensitive data processing basically two different approaches can be used, namely static and dynamic analysis, which complement each other.

\subsection{Static Analysis}
Static analysis focuses on examining the app's binary files to gain insights into its ability for retrieving and sharing sensitive information.
One approach is to analyze requested app \textbf{permissions} \cite{kollnig_are_2022,bu_ml-based_2022,verderame_reliability_2020,han_you_nodate,chang_framework_2020}, which indicates access to sensitive information like images, contacts, adID, location, or camera. Another approach is to examine \textbf{3\textsuperscript{rd} party library} integrations \cite{li_droidra_2016,kapoor_silver_2023,zimmeck_maps_2019} by matching code characteristics like class names \cite{kollnig_are_2022} or specific code features  \cite{ma_libradar_2016,backes_reliable_2016}. Identifying libraries used for e.g. advertising or telemetry can offer insights into data processing characteristics of an app.
More in-depth static analysis may involve searching for specific readable string values (e.g. URLs) \cite{abraham_mobile_2025}, or to generate call graphs \cite{li_static_2017}, which further can be used for taint analysis \cite{fu_leaksemantic_2017,alzaidi_droidrista_2020}.

Overall, however, static analyses can only indicate potential privacy violations. It could lead to false positives, e.g. if requested permissions are unused \cite{felt_android_2011}, or certain code fragments are not being executed at runtime \cite{li_static_2017}. It could also lead to false negatives as, e.g., privacy-relevant resources may still be exposed even if the app hasn't explicitly requested access permissions \cite{reardon_50_2019}.

\subsection{Dynamic Analysis}
In contrast, dynamic analyses examine the app’s actual behavior during execution. 
Thus, they provide more accurate insights into the apps behavior. However, they may lead to a higher incidence of false negatives, as certain behaviors might not be captured if specific code components are not triggered, and they require a more complex and time-consuming setup.

\paragraph{Execution Environment}
To conduct dynamic analyses an \textbf{execution environment} is required, for which either emulators \cite{cam_detecting_2019,jia_who_2019,verderame_reliability_2020} or physical devices \cite{ren_recon_2016,vanrykel_leaky_2017,xiao_lalaine_2023,kollnig_are_2022,cui_tracedroid_2022,kapoor_silver_2023} can be used. 
Thereby, emulators are more flexible and scalable, however, they can not perfectly replicate the behavior of physical devices, and apps may behave differently.
To investigate an app's overall behavior, its functionalities must be triggered through manual or automated \textbf{interaction}.
The latter offers time-saving benefits and scalability and can be achieved e.g. via UI fuzzing \cite{agrawal_survey_2022}, model-based \cite{wu_cydios_2023} 
or on-target interaction \cite{yang_appintent_2013,cam_detecting_2019}.
In contrast, manual interaction is more time-consuming, but offers greater flexibility and human behavior and thus also prevents, e.g., bot detection mechanisms.

\paragraph{Network Traffic Recording}
An important part of dynamic app analysis is the investigation of network requests, esp. regarding receiving entities and transmitted sensitive data. Network traffic of an app can be recorded remotely by redirecting it, e.g., through a proxy server \cite{kollnig_are_2022,kapoor_silver_2023}, a VPN tunnel \cite{vanrykel_leaky_2017,ren_recon_2016}, or via a Wi-Fi \ac{ap}. Another possibility is to record the traffic directly on the device using its VPN interface \cite{faranda_pcapdroid_2024} or via network traffic capturing tools\footnote{e.g. tcpdump for android: \href{https://androidtcpdump.com}{androidtcpdump.com}} \cite{sikos2020packet}.  
Packet sniffing tools as well as proxies and APs record all of the device’s traffic, resulting in possible ‘traffic mixing’ where background traffic from other apps or system services can interfere with analysis results.
To isolate app specific traffic unnecessary services has to be disabled \cite{kollnig_are_2022} or recordings has to be filtered \cite{cui_tracedroid_2022}.
VPN-based approaches, however, can leverage OS-level capabilities to directly capture only app-specific traffic.

Based on the traffic flow various metrics such as timings, volume of data transmitted, or the number of receiving end points can give first insights regarding app behavior \cite{vanrykel_leaky_2017}.
Receiver IP addresses can additional be enhanced with, e.g., WHOIS or geolocation information \cite{zilberman2024survey}. 
To identify the entity behind a specific network trace and consequently the potential data processor, domain ownership or domain category information can be retrieved from specific databases \cite{xiao_lalaine_2023}, by examining domain associated privacy polices \cite{rodriguez_roi_2023}, or by comparison with curated domain owner lists (like \cite{noauthor_duckduckgotracker-radar_2024}). 

\begin{wrapfigure}{r}{0.6\textwidth} 
    \centering
    \includegraphics[width=0.6\textwidth]{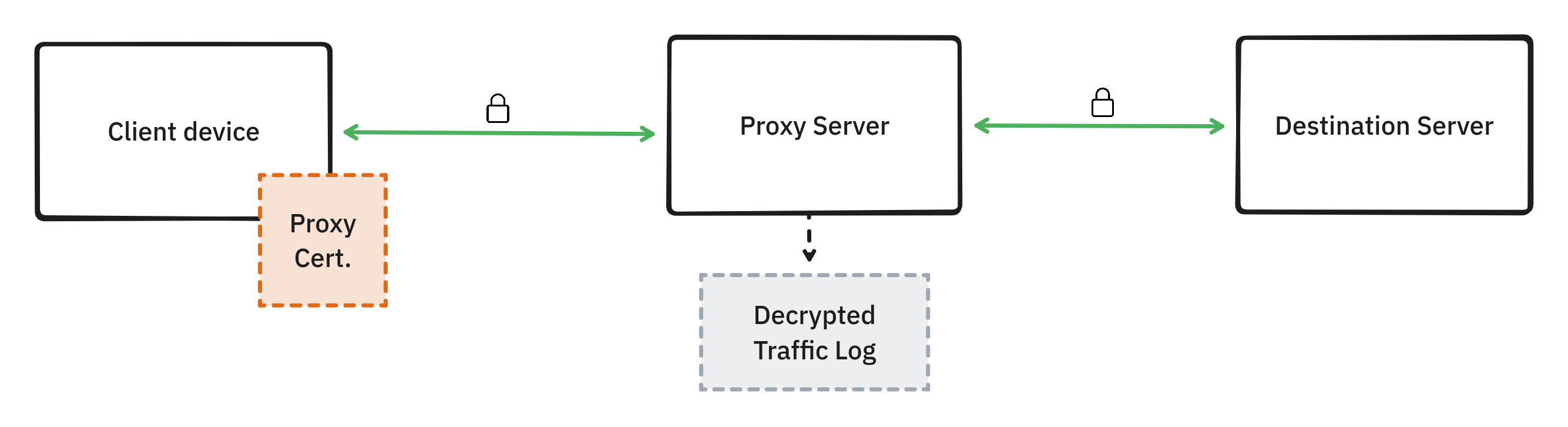} 
    \caption{Mitm Setup}
    \label{fig:mitm}
\end{wrapfigure}

\paragraph{Traffic Decryption}
To investigate the content of the traffic, \textbf{decryption} is necessary, as app traffic is commonly TLS encrypted\footnote{\href{https://security.googleblog.com/2019/12/an-update-on-android-tls-adoption.html}{security.googleblog.com/2019/12/an-update-on-android-tls-adoption.html}}.
A common approach for this is a Man in the Middle (\textbf{MITM}) approach (see Fig.~\ref{fig:mitm})  \cite{kollnig_are_2022}. Thereby a proxy server is positioned between the app and the remote server, deceiving the app into believing it is communicating with the destination server while using a custom TLS root certificate to decrypt and record the traffic, before forwarding it to the remote server. 

However, an often used countermeasure against MITM is \textit{certificate pinning}, causing the app to reject the MITM proxy’s custom certificate \cite{pradeep2022comparative}.
One approach to circumvent certificate pinning is \textbf{patching} the apps source code before conducting
the analysis, via disassembling, modifying, and reassembling, to disable certificate pinning\footnote{see e.g. \href{https://github.com/niklashigi/apk-mitm}{github.com/niklashigi/apk-mitm} or \href{https://github.com/sensepost/objection}{github.com/sensepost/objection}}. This allows traffic interception in non-rooted environments, but could be unstable due to the inherent uncertainties involved in the modifying process. Further, it compromises the integrity of the app, which could undermine the reliability of analysis results.
Another approach is called \textbf{dynamic instrumentation} \cite{luk2005pin}. 
Mobile apps typically utilize common third-party libraries. 
The functionality offered by such libraries is usually provided to the app via a fixed interface (API). Via dynamic instrumentation, API endpoints can be \textbf{hooked} during runtime, enabling the interception and/or manipulation of data flows between app and library. A comprehensive overview of various approaches and tools can be found in \cite{lopez2017survey}.
By hooking the respective \textit{certificate pinning library}, the functions that enforce a specific certificate type can be bypassed, or the custom proxy certificate can be injected \cite{noauthor_httptoolkitfrida-interception-and-unpinning_2024}.
Another, more passive way (internal logic remains unaffected), is hooking the \textit{TLS library} to extract the TLS keys, record the encrypted traffic, and decrypt it using the extracted keys afterwards \cite{fkie-cad_fritap_2024}. 
Instead of recording TLS keys, another possibility is to directly extract the clear traffic payload to be encrypted from the TLS library itself \cite{cui_tracedroid_2022}. 
With this, however, traffic sent without TLS encryption and network layers below the application layer could not be recorded. 
Overall, standard hooking scripts will fail with custom implementations or unsupported libraries, necessitating the development of scripts tailored to the app’s unique characteristics. 

\paragraph{Payload Analysis}
After decryption, network traces can be analyzed regrading sensitive data transmissions. 
The focus here is primarily on HTTP traffic, however, apps may also utilize other application layer protocols that may require distinct analysis methods.
When analyzing HTTP traffic, four key components of a network request should be examined: headers, cookies, query parameters, and the request body.
Information within could further be obfuscated, e.g. through various \textbf{encoding} algorithms, nested data structures, renamed identifiers, binary representations, or additional symmetric encryption \cite{altpeter_informed_2022}.  
A typical analysis approach involves applying a set of standard encoding algorithms, such as Base64 or URL encoding, and afterwards using \textbf{string matching} to identify known sensitive information, like common attribute names (e.g., adId, location), user-generated, or device data \cite{nguyen_share_2021,kollnig_are_2022}. 
Nguyen et al. \cite{nguyen_share_2021} proposed an approach that analyzes multiple installations of the same app, comparing recorded network traffic to identify consistently high-entropy values as unique identifiers.
The adapter based approach on the other hand uses ‘\textbf{adapters}’ designed to de-obfuscate specific, reverse engineered data transmissions to popular tracking endpoints \cite{datenanfragende_e_v_tweasel_2024,altpeter_informed_2022}. However, this approach is not easily scalable and requires continuous updates to adapters.
In addition, some studies present different \textbf{machine learning} strategies to detect Personally Identifiable Information (PII) without requiring prior knowledge of particular data points \cite{ren_recon_2016,shuba_privacy_2018}.
To enhance the accuracy of detecting sensitive payloads, multiple strategies are often combined \cite{shuba_privacy_2018,nguyen_share_2021}.

\paragraph{Local App Behavior Analyses}
Besides network traffic analysis, further dynamic methods aim to gain a deeper understanding of an app's inner workings. 
One method in this field is to record call stacks during runtime\footnote{illustrating the sequence of function calls} \cite{cui_tracedroid_2022,xiao_lalaine_2023}. By matching these to network calls, specific code segments responsible for leaking sensitive information can be identified. Subsequently, third-party libraries associated with specific network flows can be identified. Thus, conclusions can be drawn about the purpose \cite{xiao_lalaine_2023} and liability \cite{cui_tracedroid_2022} of particular data flows. 
Another approach involves recording sensitive API calls by hooking into system or library functions to monitor when apps first access sensitive data \cite{xiao_lalaine_2023,verderame_reliability_2020}. This enhances the understanding of how permissions are utilized within the app. It also allows for tracing information flows from their sources to the point of leakage.

\paragraph{Privacy Policy Analysis}
Another point of interest is the comparison of app privacy policy to its real behavior, also known as \textbf{flow-to-policy consistency}. This consists of three steps: retrieving the policy \cite{zimmeck_maps_2019}, extracting described privacy practices \cite{andow_policylint_2019,verderame_reliability_2020}
and comparing them to static and/or dynamic analysis results \cite{bui_consistency_2021,tan_ptpdroid_2023,andow_actions_2020}. 
However, due to the lack of standardized, machine-readable policies, unreliable and complex policy contents must be interpreted and compared, making this approach highly prone to errors.\\

\section{Existing Analysis Frameworks}
\label{sec:related-work}
Based on the variety of individual approaches and tools that can be used to identify privacy violations in mobile apps, frameworks exist that aim to simplify the app analysis process. These frameworks target different objectives and user groups.

\paragraph{}
\textit{PlatformControl} \cite{kollnig_platformcontrol_2024} was developed for large-scale analyses
of Android and iOS apps, and supports static as well as dynamic approaches. The latter focuses on analyzing the app's network streams on physical devices, using MITM to record and decrypt HTTPs requests thereby disabling certificate pinning. App interaction or emulated devices are not supported.
However, developed for research purposes, it lacks a centralized interface and bundled installation. Furthermore, it is no longer maintained and the utilized libraries are outdated.

\paragraph{}
\textit{Mobilsicher AppCheck} \cite{institute_for_technology_and_journalism_app-check_2024} was an initiative by a German non-profit organization that published privacy reports for Android apps aimed at everyday mobile app users. Thus, they developed tools for conducting static and network traffic analyses utilizing a simple MITM approach and automated app interactions. However, the project was recently discontinued and all AppCheck-related resources were taken offline\footnote{\href{https://mobilsicher.de/in-eigener-sache/mobilsicher-de-sagt-tschuess}{mobilsicher.de/in-eigener-sache/mobilsicher-de-sagt-tschuess}}. 

\paragraph{}
\textit{PiRogue} \cite{esther_pirogue_2025,defensive_lab_agency_docu_pirogue_2024} is a comprehensive and actively maintained forensic, mobile analysis platform. It focuses on analyzing the network behavior of Android and iOS apps on physical devices, excluding static and emulation-based methods.
For this purpose, a separate (virtual) machine (Raspberry Pi) is configured as a Wi-Fi AP. This enables PiRogue to intercept the traffic of the mobile device connected to it.
Furthermore, Pirogue employs dynamic instrumentation to capture TLS encryption keys, as well as to collect specific device ids, record cryptographic operations, and trace socket operations to identify the libraries initiating network requests. Recorded data can be exported to their web platform Colander for further investigation and report generation. However, gaining insights in Colander is an iterative and manual process. It requires users to actively engage with the data, such as by creating custom payload filters. While the combination of PiRogue and Colander provides a powerful tool for in-depth analysis, its complexity poses a significant learning curve for users, requiring manual configuration and setup across two distinct software environments. 

\paragraph{}
\textit{MobSF} \cite{abraham_mobile_2025} is another feature rich mobile analysis platform, which is actively maintained. Next to privacy analysis, its main purpose is the investigation of app security and malware. 
MobSF is shipped as a combined docker application and can be controlled using a unified web interface. Besides its in-depth static analyses, it enables traffic investigation using a MITM approach. Meanwhile, it provides instrumentation scripts, working around certificate pinning and root detection. MobSF’s support for mobile os versions is outdated and does not accommodate physical devices for analysis, limiting users’ ability to evaluate app behavior in realistic environments. Although being comprehensive, the generated reports lack aggregated findings that provide an overview of the collected data and payload investigation has to be done manually.

\paragraph{}
A distinct privacy focus is offered by \textit{Tweasel} \cite{datenanfragende_e_v_tweasel_2024}, an actively maintained analysis tool-suite. 
It is intended to enable users to identify unlawful app behavior and to submit complaints to app providers and the responsible law enforcement authorities in a largely automated manner. It solely utilizes a dynamic analysis approach based on MITM traffic recording and decryption of HTTP traffic with the ability to bypass common cert pinning libraries using dynamic instrumentation. Payload analysis is done via the adapter based approach.
Key components are divided into different modules for recording, postprocessing, and reporting, which enables reuse across different projects. Combining these modules into a comprehensive UI, however, is beyond the project's scope. Tweasel contributed to automating the installation and configuration of the required analysis tools. 

\paragraph{}
To the best of our knowledge there exist no toolkit which offers a streamlined analysis pipeline and UI specifically designed for comprehensive and holistic privacy analysis of mobile apps. Some tools are overly narrow in their analytical scope, while others offer privacy analysis as a secondary aspect, as consequence especially lacking comprehensive privacy focused reporting and guidance through the analysis process. Moreover, these tools usually require additional setup effort to operate and only support either emulated or physical devices.
Examples such as PlatformControl and Mobilsicher AppCheck illustrate that app analysis is a highly dynamic field, with analytical approaches and tools rapidly becoming outdated.

\section{Simplifying the Privacy Analysis of Apps - mopri}
Given the current state of app analysis, the following section presents an analysis suite named \textbf{mopri}, which aims to simplify and enhance the app privacy investigation process. Built around existing, well-maintained analysis tools and strategies, mopri seeks to create a cohesive software solution that eliminates the need for manual setup and guides users through the entire analysis process, from configuration to detailed reporting, within a single user interface. 
Mopri was prototypically implemented as a web application, initially focusing on the analysis of Android apps\footnote{\href{https://github.com/dfd-tud/mopri}{github.com/dfd-tud/mopri}}. The selection of modules includes key functionalities such as setting up a controlled execution environment, permission and tracking library investigations, as well as network traffic analyses, where common TLS encryption can be circumvented. Mopri's modular design allows for future enhancements and the integration of additional features as the project evolves, laying the groundwork for ongoing development in response to user needs.

\subsection{Requirements}
To achieve this, mopri was conceptualized and developed based on the following foundational requirements derived from identified gaps:

\paragraph{Robustness} Measures like error handling and the provision of various different investigation methods are required to ensure reliable and accurate results.

\paragraph{Modularity} In response to evolving challenges within the mobile app landscape, the system should follow a modular architecture that utilizes standardized data exchange formats and enables modifying, updating and replacing analysis methods. To achieve this, different components, such as the reporting or analysis pipeline, should be clearly separated. Additionally, the different analysis methods utilized in the pipeline need to be encapsulated in modules, allowing for the replacement of specific tools or the addition of entirely new investigation methods.

\paragraph{Analysis Capabilities}
At its core, the analysis pipeline should support both static and dynamic approaches to leverage the advantages of each, as outlined in Section \ref{sec:background}, with a particular emphasis on analyzing an app's communication during runtime.

\paragraph{Leverage Existing Resources}
The analyses itself should draw on well-established analysis tools and methods in order to utilize their capabilities and ensure that mopri builds on proven and up-to-date solutions.

\paragraph{Flexible Mobile Device and Interaction Choice}
Regarding the execution environment, users should have the option to use their own physical mobile device or an automatically initiated emulator. 
Furthermore, while mopri should automate the analysis process to the greatest extent possible, it shall allow users to retain manual control over interactions with the app during a dynamic analysis. This capability enables users to customize their interactions based on the specific app characteristics. 

\paragraph{User Guidance}
Mopri should streamline the initial setup and investigation process by automating installation and configuration steps, while providing a unified interface that consolidates all functionalities for a seamless user experience. Additionally, it should offer contextual help that outlines the system's background processes, available tools, and their functionalities, ensuring users remain informed and supported throughout the analysis.

\paragraph{Reporting}
Finally, all data collected by mopri's analysis modules should be used to compile a comprehensive report that outlines and summarizes the findings and insights gained throughout the analysis process for various user groups, including privacy analysts, developers, data protection authorities, and everyday phone users. The report should facilitate understanding and interpretation of the results by incorporating visualizations and enriching the content with background knowledge.

\subsection{Overview}
In response to the requirements, a concept was developed which consists of the following three integral parts, as displayed in Figure~\ref{fig:concept-stages}:

\begin{figure}
    \centering
    \includegraphics[width=\linewidth]{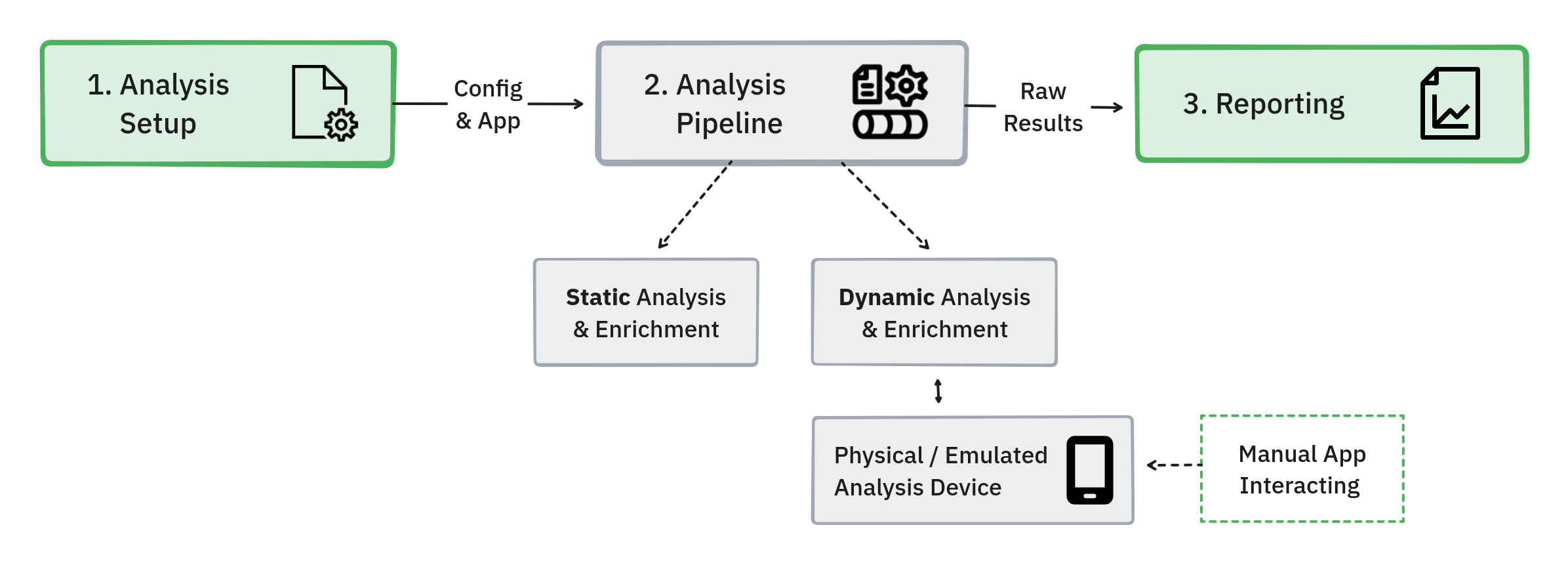}
    \caption{Conceptual Design of mopri}
    \label{fig:concept-stages}
\end{figure}

\paragraph{1. Analysis Setup:} Analysts can setup and initiate a new privacy analysis for a specific app binary using the integrated setup wizard. It allows users to choose from various analysis modules offered by mopri and configure them as needed. Assistance is provided throughout the whole process.

\paragraph{2. Analysis Pipeline (Data Collection \& Enrichment):}
Upon initiating the analysis, the chosen modules are automatically executed in the correct order via mopri's analysis pipeline. This pipeline incorporates both \textit{static and dynamic} analysis approaches, organizing the analysis steps and their corresponding tooling into interchangeable modules. The pipeline collects and records raw analysis data, such as network requests, and enriches it with relevant context to facilitate user understanding of the app's behavior and highlight privacy-critical actions. Data enrichment is a crucial step, as it enables effective reporting and aids users in evaluating the findings. 
The analysis pipeline in mopri is characterized by a high degree of automation, requiring only limited user input. The user only has to configure and start the analysis, to interact with the app to induce potentially privacy-relevant behavior during traffic recording, and finally to terminate the analysis. Throughout the process, mopri provides users with status updates and error notifications.
    
\paragraph{3. Reporting} For each analysis, the raw data and enriched results are compiled into a report tailored to the user's needs. These reports are accessible through the UI for all previously conducted analyses. The report further supports users by highlighting key findings and providing additional explanations about critical data points and the overall analysis for better comprehension. Two representation formats are available: the \textit{interactive report}, which includes interactive visuals that allow users to delve deeper into the presented findings; and the \textit{static PDF report}, which provides a concise overview of the analysis results and is therefore suitable for sharing with other parties.\\

Derived from these stages, mopri can be divided into a frontend and a backend. The frontend manages the user interface and report visualization (stage 1 and 3), while the backend handles the analysis steps (stage 2) and result storage. This separation enhances flexibility by decoupling these functions, allowing both components to communicate through a common interface (see Sec.~\ref{sec:mopri-pipeline}).

\subsection{Mopri From a User Perspective}
Besides direct interaction with the analysis device during dynamic analysis, users are intended to engage exclusively with mopri's user interface to conduct analyses and study their reports. The prototype implements the user interface as a web application using the JavaScript framework Vue \cite{you_vuejs_2024}. The prototypical interface features various views that guide users through the different stages of mobile app analysis, as follows.

\subsubsection{Setup Wizard}
To initiate a new analysis, the setup wizard guides the user through various configuration steps, which are organized and presented in a tabbed dialog.
On the first tab, metadata, e.g. title or annotations, are requested to facilitate later identification and referencing of the investigation. Furthermore, users are prompted to either upload a new app package\footnote{As of now the app needs to be manually obtained, e.g., from 3\textsuperscript{rd}-party app stores, like apkpure.com} or select one that was previously uploaded.

Afterwards, the individual analysis components can be selected and configured separately for static and dynamic analysis. If only one analysis type is needed, the other can be disabled to speed up runtime.
While static methods necessitates no additional configuration, the tab for dynamic analysis presents an array of configuration options. First, the user has to decide whether to use a physical or emulated device. If the latter is selected, additional configuration options become available to setup the emulator or choose an existing one, that is already installed on the system.
Next, a traffic recording method has to be selected from the list of available approaches, each of which uses a different approach and tools to capture and decrypt an app's network traffic (see Sec.~\ref{sec:mopri-pipeline}). To assist users in their selection, explanatory notes are provided for each method.

\begin{figure}
    \centering
    \includegraphics[width=\linewidth]{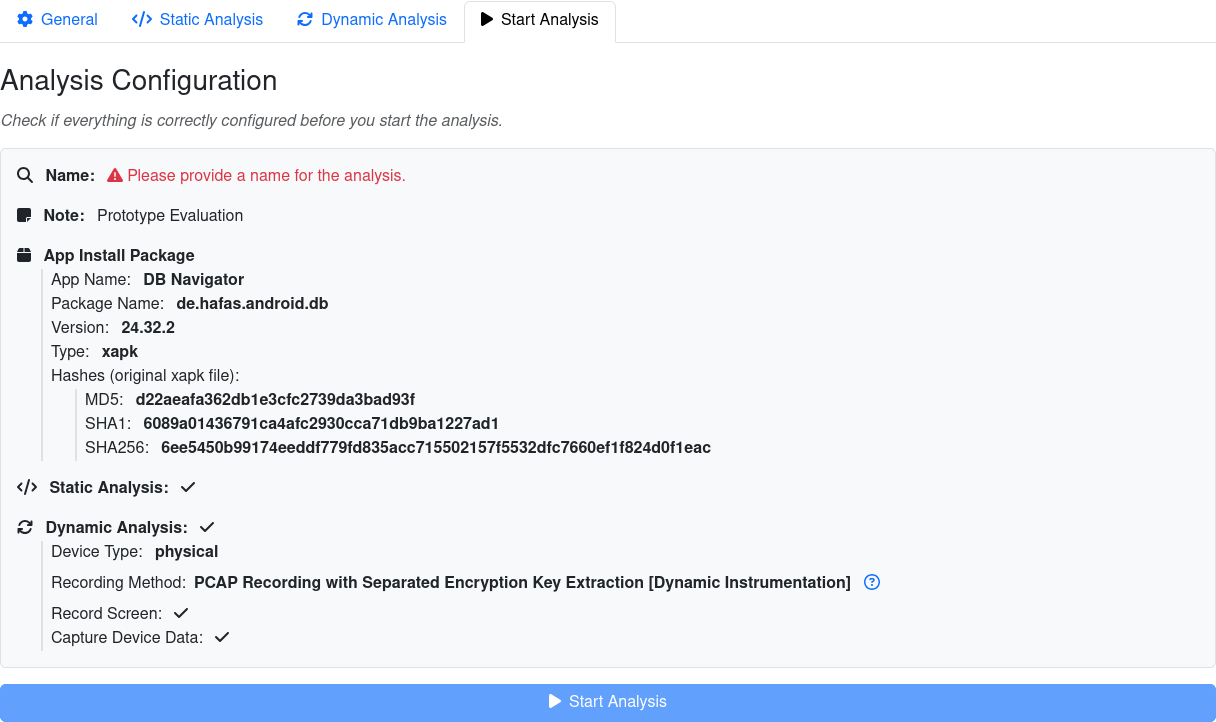}
    \caption{Analysis Summary within the Setup Wizard}    
    \label{fig:ui-wizard-summary}
\end{figure}

At the end of the setup process, users receive an overview of the generated configuration before starting the analysis (see Fig.~\ref{fig:ui-wizard-summary}).

\subsubsection{During Analysis}
During the execution of the analysis, users are provided with detailed log information. The analysis procedure is initiated in a sequential manner, wherein the static analysis is conducted first, without requiring any user interaction, followed by the dynamic analysis.
Before starting the latter, mopri ensures it has access to the correct Android device. 
If a physical device is selected, the user is prompted to connect a rooted Android device to the host machine via USB. Conversely, if an emulator is chosen, it is launched by the Android emulation software in a separate window on the host machine, allowing the user to interact seamlessly with the emulated device.
After the execution environment is successfully initialized, an automated setup process starts, installing and configuring the required analysis tools and the app to be analyzed, all without any additional user interaction.

Once the setup is complete, the recording process is initiated, and the app is automatically launched on the selected device. The user is then notified to interact with the app. This step is entirely user-driven, with the system recording the required data in the background without activating any additional processes on the device during interaction.  Throughout this phase, a termination button is visible, allowing users to end the analysis at any time.

Upon termination, the device is restored to its pre-analysis state by removing all installed tools. Subsequent enrichment steps are then executed, and a link to the corresponding report is displayed on the screen, marking the analysis as complete.

\subsubsection{Reporting}
As part of the user interface an \textbf{interactive report} is compiled for each analysis, based on the data provided by the analysis pipeline. The report is organized in the following tabs:

\paragraph{About} The first tab presents details about the test configuration and the analyzed app, along with technical information regarding the device used for dynamic analysis. Overall, this information was added to aid traceability and reproducibility of the investigation.

\begin{figure}
    \centering    
    \begin{minipage}{0.4\linewidth} 
        \centering
        \includegraphics[width=\linewidth]{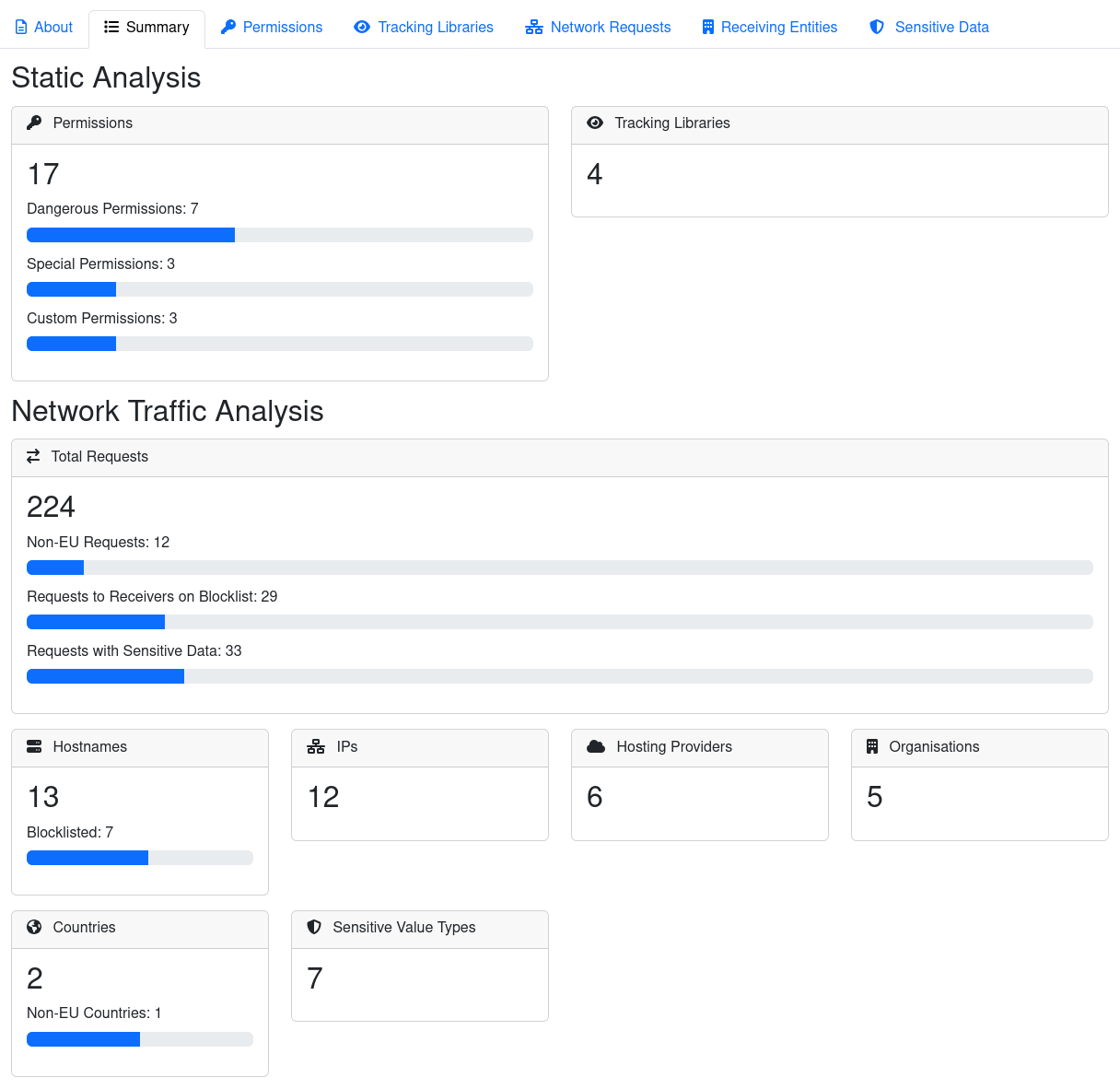}
        \caption{Aggregated Analysis Results}
        \label{fig:ui-report-summary}
    \end{minipage}%
    \hfill 
    \begin{minipage}{0.57\linewidth} 
        \centering
        \includegraphics[width=\linewidth]{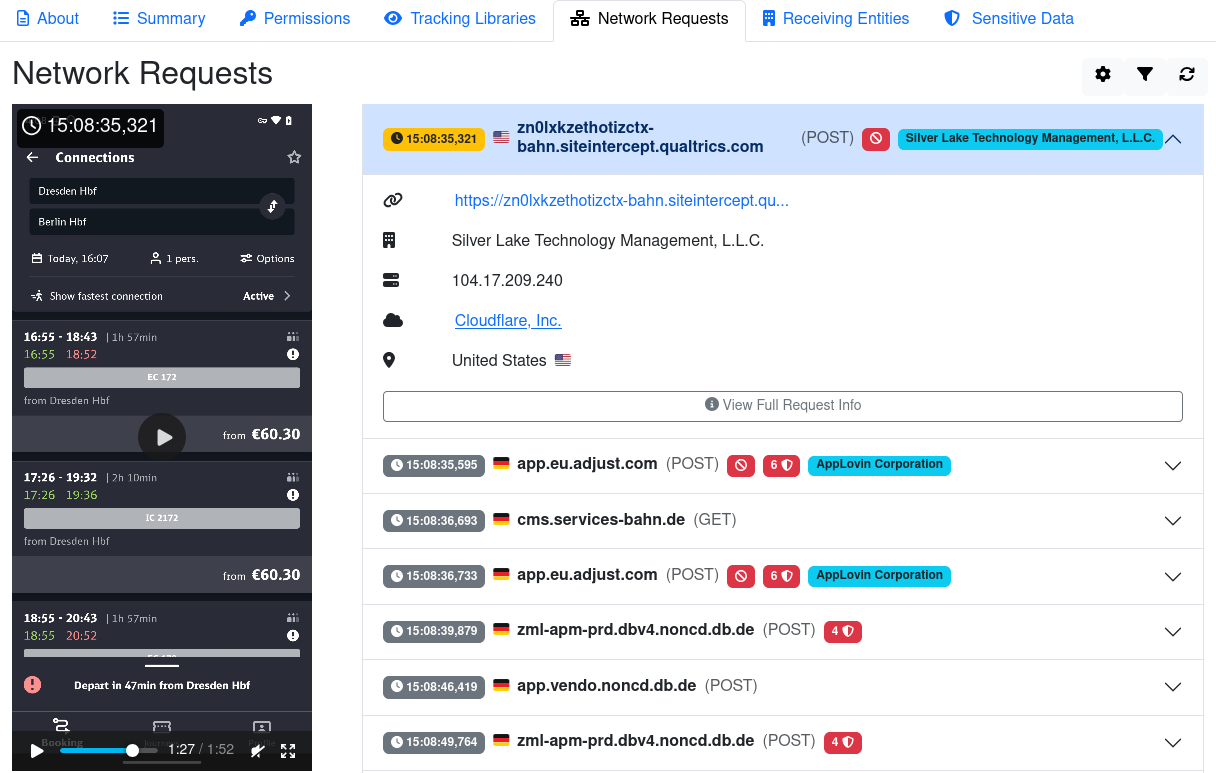}
        \caption{Network Requests}
        \label{fig:ui-report-requests}
    \end{minipage}
\end{figure}
    
\paragraph{Summary}
As illustrated in Figure~\ref{fig:ui-report-summary}, the summary tab provides a concise overview of the results to facilitate comparison between analyses. This is done through aggregated metrics, like the number of permissions identified, recorded network requests and entities that received data.
    
\paragraph{Permissions} This tab lists all permissions extracted from the app binaries, along with corresponding descriptions and highlights sensitive permissions. 
    
\paragraph{Tracking Libraries}
Similar to the permissions overview, here details regarding third-party tracking libraries found within the app binaries are shown. Each entry is enriched with additional information, including its purpose and the company distributing it. 
    
\paragraph{Network Requests}
Presenting the results of a dynamic analysis this tab provides an comprehensive insight into the recorded and decrypted HTTP requests (see Fig.~\ref{fig:ui-report-requests}). Each request is displayed as a list entry with suitable iconography, containing enriched metadata and decrypted payloads, which are outlined in a pop-up window upon request. To relate the list of recorded requests to the interactions with the app, the video playback of the recorded app screen during analysis (left) is synchronized with the highlighted entries in the request list (right).
    
\paragraph{Receiving Entities}
This tab provides an overview of the receiving entities in the form of traffic endpoints identified in the recorded network requests. It is enriched with corresponding company and location data, as well as aggregated metrics such as the number of requests directed to each specific endpoint.

\paragraph{Sensitive Data}
Concluding the report, the last tab compiles sensitive data found within the payloads of recorded network requests.\\

In addition to the interactive report, a static report can be generated that removes interactive elements and presents the included data points adopted to a printable format.

\subsection{Mopri From a Technical Perspective - The Analysis Pipeline}
\label{sec:mopri-pipeline}
\begin{figure}
    \centering
    \includegraphics[width=\linewidth]{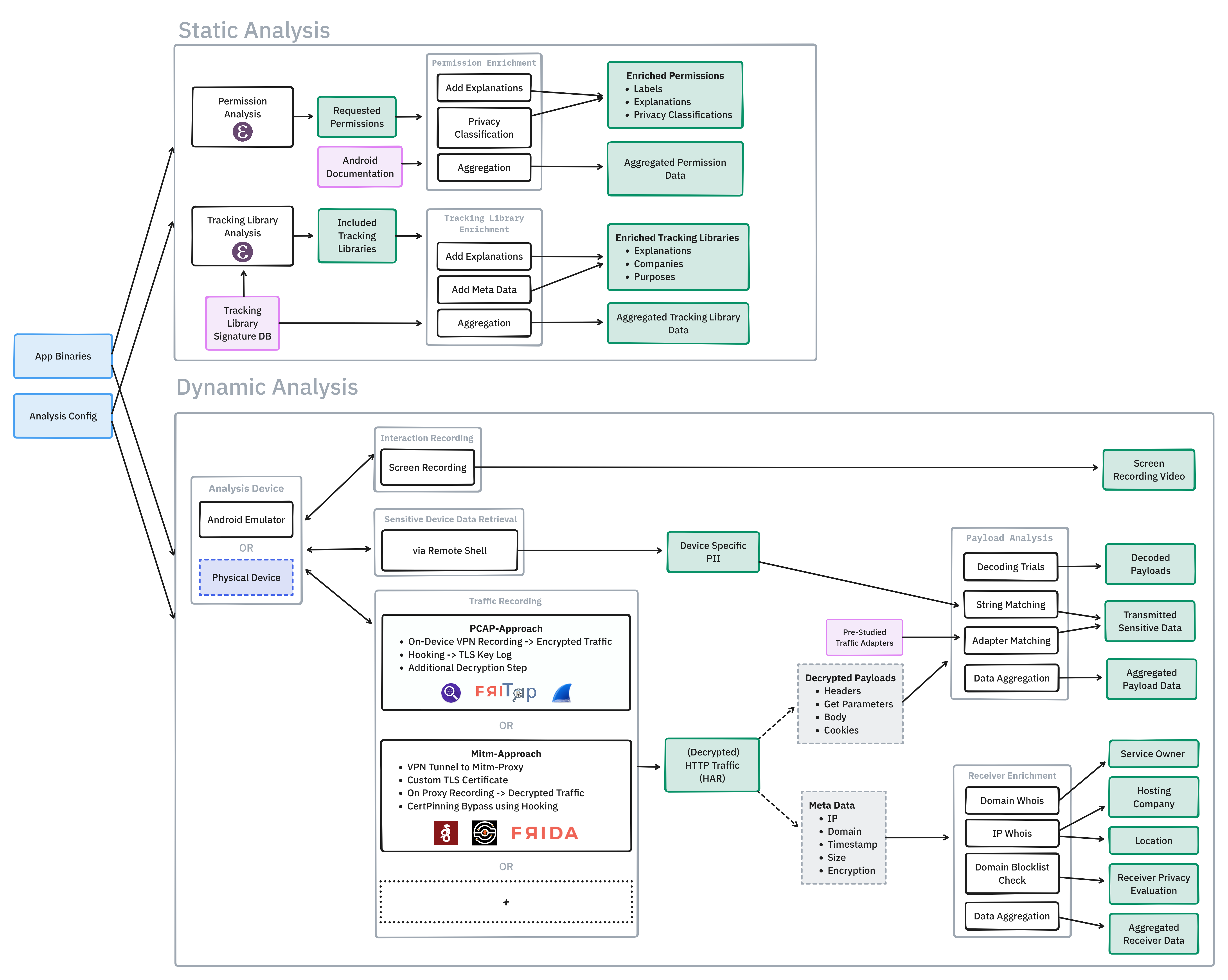}
    \caption{Analysis Pipeline}    
    \label{fig:mopri-analysis-pipeline}
\end{figure}

The heart of mopri is its modular analysis pipeline. It is responsible for gathering and enhancing the analysis data (stage 2), as shown in Figure~\ref{fig:mopri-analysis-pipeline}.  As part of the prototype, the analysis pipeline is integrated within the backend, which also persists the collected analysis data and uploaded app binaries and provides them to the user interface upon request. The prototype supports the analysis of Android applications provided in APK\footnote{\href{https://developer.android.com/guide/components/fundamentals}{developer.android.com/guide/components/fundamentals}
} or XAPK\footnote{\href{https://apkpure.com/xapk.html}{apkpure.com/xapk.html}}. The prototype's backend is developed as a Node.js application. 
To facilitate data exchange and control flow with the fronted, the backend exposes a centralized API following the REST paradigm.

The analysis pipeline is designed following a modular architecture, encapsulating individual analysis and enrichment steps, along with their requisite tools, into distinct modules. These modules are interconnected through orchestration components. A primary orchestration module acts as the entry point to the pipeline, triggered by the corresponding API endpoint. It manages the overall analysis process and monitors ongoing analyses.
Mopri integrates both static and dynamic approaches (see Sec.~\ref{sec:background}). Consequently, the pipeline diverges into these two distinct types, each orchestrated independently and executed sequentially. This separation is necessitated by the differing control sequences: static analysis can be conducted independent from the user, whereas dynamic analysis require user interaction with the device and relies on the user to determine when to finish the analysis.
The analysis configuration, compiled during the analysis setup, is transmitted via the API in the form of a single file and propagated throughout the analysis pipeline. This configuration dictates which analysis modules need to be initiated by the orchestration modules and provides the necessary configuration flags for each analysis module. 

Each module is responsible for storing its outputs using a centralized storage module, which utilizes a file based storage system. These outputs can subsequently be accessed by the UI through the API to generate reports. To ensure consistency, schemes describing the outputs of each module are maintained across both the backend and frontend. When a module is replaced, its outputs must conform to the same scheme as the previous module. Further, each analysis module adheres to standardized interfaces defining the routines exposed by a module. This simplifies replacing individual modules.

\subsubsection{Static Analysis Modules}
For static analysis the corresponding module utilizes tools provided by the Exodus Privacy Project \cite{exodus_privacy_exodus_2024}, to extract requested permissions and included tracking libraries from an app's binaries. In a separated step these data points are enriched with user-friendly labels and descriptions, highlighting privacy-relevant permissions and categorizing tracking libraries by their purpose and distributor, utilizing the Android's permission documentation\footnote{\href{https://developer.android.com/guide/topics/permissions/overview}{developer.android.com/guide/topics/permissions/overview}} and Exodus' tracker database. 

\subsubsection{Dynamic Analysis Modules}
Regarding dynamic analysis a separate parent module orchestrates a diverse array of sub-modules, many of which operate concurrently until they receive the signal to stop passed through from the user interface. For each dynamic analysis, the required tools are automatically installed on the analysis device and removed after the analysis is finished. 
After the initial setup and once the analysis begins, a signal is sent to the UI to prompt the user to interact with the app to trigger relevant app traffic.
Conversely, users can signal the pipeline at any time to stop the ongoing analysis and perform subsequent cleanup and post-processing steps.
Overall, depending on the user configuration, a dynamic analysis process generally encompasses the following steps:

\paragraph{Starting}
    \begin{enumerate}
        \item[1] Ensuring an active analysis device.
        \item[2] Extracting sensitive device data required for payload analysis.
        \item[3] The selected network \textit{traffic recording} module is initiated through a factory method, and the corresponding \textit{device preparation} routine is executed.        
        \item[4] The \textit{screen recording} module is initiated and the video recording started.        
        \item[5] The network traffic recording is started.
    \end{enumerate}

\paragraph{Stopping}
    \begin{enumerate}
        \item[1] The \textit{stop routines} of the running analysis modules are called in the reverse order of their initiation. To facilitate this, the analysis modules are stored in a list according to the order in which they were started, allowing for the flexible addition of further parallel-running analysis modules. For instance, the screen recording must be initiated before and concluded after the network traffic recording to ensure that all application behavior during the recorded traffic is captured.
        \item[2] The respective \textit{cleanup routines} are executed in the same reverse order as the stop routines.
        \item[3] \textit{Post-processing routines} are invoked in reverse order to transform module outputs into the required format.
        \item[4] If applicable, the emulator is stopped.
        \item[5] The parent orchestration module is signaled about the end of the dynamic analysis in order to continue with further analysis steps, in this case to end the analysis and notify the front end accordingly.
  \end{enumerate}

\subsubsection{Execution Environment}
In line with the requirements, the analysis pipeline supports two types of analysis devices: a rooted physical device connected to the host machine, or an emulator based on an Android Virtual Device (AVD) configuration. If an emulator is selected, it is launched using the specified name of an existing AVD. 
The logic for setting up AVDs, which includes downloading the corresponding system images, is implemented outside the analysis pipeline based on the Andromatic\footnote{\href{https://github.com/tweaselORG/andromatic}{github.com/tweaselORG/andromatic}} library and is triggered during the analysis setup if requested by the user.
The emulator control logic, as part of the dynamic analysis pipeline, is encapsulated in a dedicated module, which exposes start and stop routines to the orchestration module. The prototype utilizes the Appstraction\footnote{\href{https://github.com/tweaselORG/appstraction}{github.com/tweaselORG/appstraction}} library to instrument Android's emulation software accordingly.
Further procedures for controlling the Android emulator can be used by replacing the emulator module.
In contrast, when a physical device is selected for analysis, no separate device module is loaded. The orchestration module ensures that only one device is connected at a time (either the emulator or a single physical device) before proceeding with the analysis. This is due to a limitation of Appstraction, which cannot differentiate between multiple connected devices and is used across the pipeline for communicating with the Android device via the Android Debug Bridge (ADB). 

\subsubsection{Visual App Recording}
To ensure traceability of analysts' app interactions and enable correlating the app's visual behavior with the recorded network requests, mopri integrates an additional interaction recording module. This module uses Android's builtin screen recording tool to capture the device's screen during analysis (see Fig.~\ref{fig:ui-report-requests}), invoked trough ADB. The resulting video file is automatically pulled from the analysis device and saved along the other dynamic analysis results.

\subsubsection{Network Traffic Recording}
To facilitate robustness against various challenges within the mobile ecosystem, like apps using diverse encryption libraries and security measurements, mopri offers multiple approaches for recording and decrypting network traffic. 
Users are able to select a suitable approach that fits the circumstances of the app they want to analyze. For the prototype, based on the findings in Section~\ref{sec:background}, these approaches are centered around two well-established methodologies that have proven effective and are supported by reliable tools.

\paragraph{MITM Recording (with Certificate Pinning Bypass)}
The first module utilizes Cyanoacrylate\footnote{\href{https://github.com/tweaselORG/cyanoacrylate}{github.com/tweaselORG/cyanoacrylate}}, which is part of the Tweasel tool suite (see Sec.~\ref{sec:related-work}). This module redirects the app's traffic through a mitmproxy \cite{cortesi_mitmproxy_2024} server utilizing the WireGuard\footnote{\href{https://github.com/WireGuard/wireguard-android}{github.com/WireGuard/wireguard-android}} VPN client and a custom TLS certificate. It optionally bypasses certificate pinning with the help of Frida\footnote{\href{https://frida.re}{frida.re}} and a set of hooking scripts \cite{noauthor_httptoolkitfrida-interception-and-unpinning_2024}. The decrypted traffic is recorded by the proxy in HAR \cite{odvarko_http_2012} format. Due to its widespread compatibility and ease of processing, this format serves as the primary output format for decrypted network traffic within the prototype.

\paragraph{On-Device VPN Traffic Recording (with Separate TLS Key Extraction)}
The second module employs the VPN client PCAPDroid \cite{faranda_pcapdroid_2024} to capture raw traffic directly on the mobile device. For subsequent decryption, TLS keys are extracted separately from common TLS libraries using FriTap \cite{fkie-cad_fritap_2024}. 
This method enables data traffic to be recorded even in cases where decryption failed, so that metadata can still be analyzed. Moreover, unlike the previous HAR-based method, this method is able to extract recipients of non-HTTP traffic as it records in the packet-based PCAP \cite{harris_pcap_2024} format as an intermediate step. However, tests with example apps from the German app market revealed that, in some cases, the mitmproxy approach successfully decrypted traffic when FriTap was unable to record the TLS keys. Consequently, both modules have their own utility.

\paragraph{}
In summary, this results in four distinct recording methods (mitm with/without cert pinning bypass; on-device recording with/without TLS key extraction) that users can select based on the app's specific circumstances. While the existing analysis approaches enable mopri to cover a wide range of applications that utilize standard encryption and communication libraries, customized security measures may still necessitate manual analysis. Such cases cannot be addressed by mopri unless a custom analysis module is developed to tackle these specific challenges.

The pipeline's underlying software architecture enables the integration of new recording modules as they become available. Each recording module encapsulates the necessary tooling and specific routines. Thereby common sub-tasks such as installing or running a VPN client are provided as separated modules that can be shared across different recording approaches. All network traffic recording modules adhere to a common interface that defines standardized routines, namely setup, start, stop and cleanup, so that the orchestration module can control the traffic recording life-cycle regardless of the selected module.
Additionally, each recording module needs to store the recorded requests in the common HAR schema, that is used by subsequent analysis and recording modules. Thus, if necessary, they may implement a post-processing routine to convert the recorded requests into HAR, as is done for on-device recording using PCAPdroid. During a dynamic analysis, the required recording module is initialized based on a unique key assigned to each recording approach. The selected key is transmitted as part of the analysis configuration.

\subsubsection{Network Traffic Enrichment}
\label{p-enrichment}
As part of the analysis, the recorded traffic data is enriched with additional information, consisting of two main components: traffic metadata enrichment and payload analysis.

\paragraph{Metadata enrichment} focuses on data-receiving services identified by their IP addresses and corresponding domains within the raw traffic records.
If a PCAP file is available (as e.g. produced by PCAPdroid) the command line tool tshark\footnote{\href{https://tshark.dev}{tshark.dev}} is used to extract further entities based on TLS handshakes. This is done to cover requests that are missing in the HAR file used for the analysis because they could not be decrypted or utilize an application protocol other than HTTP.
The location of each receiving server and corresponding hosting company are identified based on the IP, utilizing the online service IPWhois\footnote{\href{https://ipwhois.io}{ipwhois.io}}. The company potentially offering the identified service is determined by matching the domain with DuckDuckGo's TrackerRadar \cite{noauthor_duckduckgotracker-radar_2024}, to find the most relevant tracking related entities. Additionally, a first attempt is made to classify the service's privacy impact by cross-referencing its domain with a set of privacy block lists, as used by DNS-based network filters and advertisement blockers\footnote{e.g. \href{https://easylist.to}{easylist.to}}.

\paragraph{Payload analysis} employs a generalized approach to counter common content encoding strategies, esp. URL, JSON (including recursive multi layered encoding), Base64 and GZIP encoding, aiming to produce a human-readable representation of the content. After decoding, regex patterns derived from a set of sensitive fixed device data are matched with the payloads to reveal transmitted sensitive information. In the context of the prototype, this analysis is based solely on a small selection of data points extracted from the analytics device during dynamic analysis, including its advertising ID and other lower entropy device attributes such as model number, chip architecture and manufacturer, which are regularly transmitted by apps. To obtain more meaningful results, the payload analysis was extended by Tweasel's adapter approach\footnote{\href{https://github.com/tweaselORG/TrackHAR}{github.com/tweaselORG/TrackHAR}}, which extracts detailed data for a limited number of known tracking endpoints.\\

Alongside the enrichment, all collected meta and payload data is aggregated to provide insights for the corresponding summary screen within the report, including metrics like the overall number of requests, the number of contacted domains, and the amount of sensitive information found.

\section{Outlook and Open Challenges} 
The presented prototype of mopri already puts many aspects of the underlying concept into practice and enables detailed privacy analyses of selected Android apps. However, it remains a prototype with potential for further improvements and unresolved challenges. These need to be adressed for so that users can effectively integrate mopri into their privacy analysis workflows.

One point is the integration of support for analyzing iOS apps. After Android, iOS has the second largest market share among operating systems \cite{szczepanski2018european}. At same time, it is underrepresented in mobile privacy research \cite{kollnig_are_2022}.
While some of the tools used in the prototype, such as the analysis modules of Tweasel, already provide support for iOS, it is necessary to find alternatives for components that do not. This includes the static parts based on Exodus or the emulation environment.
Furthermore, privacy policy investigation as an integral part of app analysis should be added as a new module to the analysis pipeline. Linking with other analysis results would provide a more comprehensive overview of an app's data protection practices and the associated transparency.

Existing analysis modules provide room for improvement as well. This involves adding more ways for users to provide apps, expanding analysis capabilities for non-HTTP traffic, and improving payload analysis that goes beyond string matching and considers a larger set of sensitive data, e.g., complemented by user-provided information entered during interaction with the app.
Regarding the dynamic analysis, currently, users need to interact directly with the (emulated) device itself. However, prospectively, users should be able to interact with both emulators and physical devices directly from within the UI.
Other examples include the provision of more comprehensive whois information about traffic recipients and a refined privacy classification of app permissions. This goes in hand with further refining the pipelines' software architecture and documentation to simplify the process of adding these features.
To evaluate and improve mopri's analysis capabilities and overall robustness, it needs to be systematically tested. This also includes studying differences in app behavior on emulated and physical devices, to be taken into account during analysis.
Beyond these functional improvements, mopri's UX and UI should be further improved based on a user study. This includes adding further guidance through the application and analysis process, providing different reporting templates depending on the users needs, as well as improving the data visualization in the report.

\section{Conclusion}
With mopri, we presented a new privacy analysis framework, which provides a comprehensive, adaptable, and user-centered approach to investigate privacy violations of mobile apps. Identified through a review of existing frameworks, mopri addresses the limitations of current tools by integrating key static and dynamic analyses into a modular pipeline. This integration allows for a thorough examination of app behavior, focusing on the actual leakage of sensitive data via network requests. It automates many otherwise time-consuming tasks, such as installing and setting up the necessary tools, data processing and reporting, but leaves the crucial part of interacting with the app to the users. A single user interface facilitates the analysis process, enabling users to setup their analysis, monitor it during execution and investigate provided analysis reports. While the initial evaluation highlights areas for improvement, including robustness and further functionality like supporting iOS apps, mopri establishes a solid foundation for future enhancements. As mobile apps evolve, ongoing updates and user contributions to mopri become essential to adapt to new privacy challenges, ultimately promoting accountability of app providers and improving user privacy.

\small
\bibliographystyle{splncs04}
\bibliography{ref}

\end{document}